\documentclass[useAMS,usenatbib]{mn2e}
\usepackage{graphicx,amsmath,amssymb,subfigure,rotating,epsfig}

\def\simgt{\mathrel{\lower0.6ex\hbox{$\buildrel {\textstyle >} \over {\scriptstyle \sim}$}}}
\def\simlt{\mathrel{\lower0.6ex\hbox{$\buildrel {\textstyle <} \over {\scriptstyle \sim}$}}}
\newcommand{\Msolar}{\mbox{\,$\rm M_{\odot}$}}          % solar mass
\newcommand{\mg}{Mg{\sc ii}\,\,}

\newcommand{\gtsim}{\mbox{{\raisebox{-0.4ex}{$\stackrel{>}{{\scriptstyle\sim}}$}}}}

\newcommand{\mnras}{MNRAS}

\newcommand{\apss}{A\&SS}

\begin{document}

\title[The environments of $z\sim 1$ Active Galactic Nuclei at $3.6~\mu m$]{The
  environments of $z\sim 1$ Active Galactic Nuclei at $3.6~\mu m$} 

\author[J. T. Falder et al.~2009] {J.\ T.\ Falder,$^{1}$\thanks{Email:
    J.T.Falder@herts.ac.uk} J.\ A.\ Stevens,$^{1}$
  Matt\ J.\ Jarvis,$^{1}$ M.\ J.\ Hardcastle,$^{1}$ M.\ Lacy,$^{2}$ \and
  R.\ J.\ McLure,$^{3}$ E.\ Hatziminaoglou,$^{4}$ M.\ J.\ Page$^{5}$ and G.\ T.\ Richards$^{6}$ \\ 
\footnotesize $^{1}$Centre for Astrophysics
  Research, University of Hertfordshire, College Lane, Herts, AL10
  9AB\\ $^{2}$Spitzer Science Center, California Institute of Technology, MC
  220-6, Pasadena, CA 91125, USA\\ $^{3}$
  SUPA\thanks{Scottish Universities Physics Alliance}, Institute for Astronomy, University of
  Edinburgh, Royal Observatory, Blackford Hill, Edinburgh, EH9
  3HJ\\ $^{4}$European Southern Observatory, Karl-Schwarzschild-Str. 2, 85748
  Garching bei Munchen, Germany\\ $^{5}$Mullard Space Science Laboratory,
  Holmbury St. Mary, Dorking, Surrey, RH5 6NT\\ $^{6}$Department of Physics, Drexel University, Philadelphia, PA 19104, USA}

\maketitle

\begin{abstract} 
We present an analysis of a large sample of AGN environments at $z\sim 1$ using
stacked {\em Spitzer\/} data at 3.6~$\mu$m. The sample contains type-1 and
type-2 AGN in the form of quasars and radio galaxies, and spans a large range
in both optical and radio luminosity. We find, on average, that 2 to 3 massive
galaxies containing a substantial evolved stellar population lie within a
$200-300$~kpc radius of the AGN, constituting a $>8$-$\sigma$ excess relative
to the field. Secondly, we find evidence for the environmental source density
to increase with the radio luminosity of AGN, but not with black-hole
mass. This is shown first by dividing the AGN into their classical AGN types,
where we see more significant over-densities in the fields of the radio-loud
AGN. If instead we dispense with the classical AGN definitions, we find that
the source over-density as a function of radio luminosity for all our AGN
exhibits a positive correlation. One interpretation of this result is that the
Mpc-scale environment is in some way influencing the radio emission that we
observe from AGN. This could be explained by the confinement of radio jets in
dense environments leading to enhanced radio emission or, alternatively, may be
linked to more rapid black-hole spin brought on by galaxy
mergers.  \end{abstract}


\section{Introduction}

It is now widely accepted that high-luminosity Active Galactic Nuclei (AGN)
harbour accreting super-massive black holes implying that their host galaxies
are amongst the most massive objects in existence at their respective
epochs. Indeed, many studies have now shown that AGN preferentially reside
within fields containing over-densities of galaxies
(e.g. \citealt{Hall98,Best03,Wold03,Hutchings09}). Together these points
support the idea that AGN can be utilized as signposts to extreme regions of
the dark matter density and thus the most massive dark matter haloes
(e.g. \citealt{Pentericci00,Ivison00,Stevens03}) at any given epoch. Combining
this technique with large multiwavelength surveys, like the Sloan Digital Sky
Survey (SDSS; \citealt{Adelman06}) which has identified up to 100000 broad-line
quasi-stellar objects (hereafter quasars) up to the highest measured redshifts
(i.e. z=6.4, \citealt{Fan03}), has opened up a new era in AGN research.

Many authors have addressed the question of whether the environments of
radio-loud AGN, such as radio-loud quasars (RLQs) or radio galaxies (RGs), are
any different from those of radio-quiet AGN, such as radio-quiet quasars
(RQQs), with conflicting results. The first work that compared directly the
environments of RLQs and RQQs was \cite{Yee84}, in which a marginally more
significant over-density was detected around the RLQs in their sample of
objects at $0.05 < z < 0.55$. However, a later improved study with more data
and refined techniques removed the significance of this result
\citep{Yee87}. More work on the topic was conducted by \cite{Ellingson91} who
added more faint RQQs to the \cite{Yee87} sample. As a result they reported a
significant difference in the environments preferred by RLQs and RQQs, with
RQQs in general preferring poorer environments at the 99~per cent confidence
level. At $0.9 < z < 4.2$ \cite{Hutchings99} found that RLQs occupied more
dense environments in the near-infrared than RQQs. In contrast, both
\cite{Wold01} and \cite{McLure01} found the environments of RLQs and RQQs to be
indistinguishable at z$\sim$0.2 and at $0.5 < z < 0.8$ respectively. More
recently \cite{KauffmannHeckman&Best} found that radio-loud AGN reside in
environments a factor of $2-3$ more dense than radio-quiet AGN in a large
matched sample of SDSS emission-line AGN in the local universe. These results
present us with a very mixed picture. However, it is probable that many of them
suffer in some way from small number statistics and/or significant selection
effects. Furthermore, it may be crucial to understand how the AGN and their
environments are linked at all redshifts given that large-scale radio-jet
activity could enhance (\citealt{Wiita04}) or truncate
(\citealt{Rawlings&Jarvis04}) star formation on the Mpc scale, and hence be a
crucial ingredient in semi-analytic models (e.g. \citealt{Croton06};
\citealt{Bower06}).

It is still not clear exactly what triggers the radio emission we see in
radio-loud AGN, and why we do not see similar emission in otherwise comparable
radio-quiet AGN. \cite{McLure&Jarvis04} showed that, on average, RLQs have
45~per cent more massive black holes than RQQs, and that most RLQs have a
black-hole mass greater than $10^8$\Msolar. These results suggest that in order
to be radio-loud an AGN requires a certain mass of black hole. However, at any
given black-hole mass the range of radio luminosities spans several orders of
magnitude and there are RQQs which have black holes equally as massive as the
RLQs. Thus black-hole mass cannot be the only factor that determines radio
power; perhaps the environment of the AGN also contributes.

It is also still not fully understood whether there is a true AGN radio power
dichotomy (e.g. \citealt{Ivezic}; \citealt{Lacy01}; and more recently
\citealt{White07}; \citealt{ZamfirSulentic&Marziani}), and whether the
different types of AGN are intrinsically different or whether we just happen to
be observing some whilst they are going through a period of
radio-loudness. Studying the cluster-scale environments of these objects
provides a good means of investigating this issue since, if they are identical
objects going through different phases then they should have similar
environments on these scales. Alternatively, it may be the case that the
environment is in some way linked to the differences that we see in the radio
properties of AGN.

  In this paper we concentrate on a number density analysis of a
  sample of 173 AGN fields at the single cosmic epoch of $z\sim1$,
  splitting the AGN into their classical types (RLQs, RQQs and RGs),
  as well as looking for trends with radio luminosity and black hole
  mass. We use 3.6~$\mu$m observations allowing us to sample the peak
  of the rest-frame stellar spectrum thus maximising our sensitivity
  to stellar mass. This work presents an analysis of the environments
  of the largest, most uniformly selected sample of luminous AGN yet
  assembled at high redshift. In so doing it forms an extension to
  previous studies of the environments of AGN with lower radio
  luminosities and at lower redshifts ($z<0.3$) performed with SDSS data
  (e.g. \citealt{BKHI05}; \citealt{KauffmannHeckman&Best}).

In Section 2 we give details of our data, in Section 3 we discuss the source
extraction, in Section 4 we explain the method of analysis, then in Section 5
we present our results followed by a discussion in Section 6 and a summary of
the main conclusions in Section 7. Throughout this paper we have assumed a flat
cosmology with H$_0=72$ km\ s$^{-1}$\ Mpc$^{-1}$, $\Omega_{\rm{m}}=0.3$ and
$\Omega_{\Lambda}=0.7$. All magnitudes are quoted in the AB system.

\begin{figure}
\centering
\includegraphics[width=0.95\columnwidth]{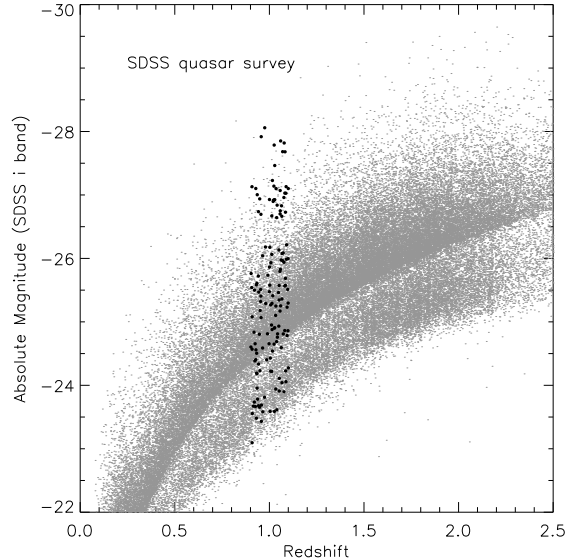}
\caption{Redshift vs optical absolute magnitude (SDSS i band) for quasars from
  the fifth data release of the SDSS quasar survey \citep{Schneider05}. The
  quasars in bold are those used in our sample, clearly showing that we span
  the 5 magnitude range in optical luminosity available at $z\sim 1$.}
\label{fig:SDSS}
\end{figure}

\section{Data}

The data presented in this paper consist of infrared images of 173 AGN taken at
3.6~$\mu$m with the IRAC camera on-board the {\em Spitzer Space Telescope}. The
sample is split into three sub-samples, all at the single cosmic epoch of $0.9
< z < 1.1$: 75 RLQs, 71 RQQs and 27 RGs. This redshift was chosen as it is the
minimum at which there is a large enough population of high-luminosity quasars
to allow comparison with the bright quasars that are observed at higher
redshifts. At this redshift the SDSS allows us to sample over 5 magnitudes in
quasar optical luminosity (see Fig.~\ref{fig:SDSS}). This sample thus enables
us to decouple luminosity generated effects from evolutionary ones, something
which has plagued many other flux density limited studies in this area. In
addition, the targets were chosen to have redshifts optimized, within the
chosen range for follow-up CO surveys with interferometers such as the Atacama
Large Millimeter Array (ALMA).  Observing both unobscured (type-1) AGN, in the
form of quasars, and obscured (type-2) AGN, the RGs, allows us to test AGN
unification schemes (e.g. \citealt{Antonucci93}).

Full details of the quasars will be presented elsewhere (Jarvis et al., in
prep) while a list of the RGs giving their main properties is given in Table
\ref{tab:RGs}. The RGs properties are taken from NASA Extragalactic Database
(NED), except for the 6C objects where the redshifts are taken from
\cite{Best96}, \cite{Rawlings01} and \cite{Inskip05}, and the 6C* and TOOT
objects which are described by \cite{Jarvis01} and \cite{Vardoulaki09}
respectively. Further details of the RGs will be presented elsewhere (Fernandes
et al., in prep).

\begin{table}
\centering
\caption{The radio galaxies used in this paper. Column 2 gives the observed-frame
  325~MHz flux density, column 3 the spectral index and column 4 the
  redshift. The 325~MHz flux densities and spectral indices
  ($S_{\nu}\propto\nu^{-\alpha}$) are calculated by fitting a power-law through
  available flux density measurements taken from the NASA Extragalactic
  Database (NED), except for 6C* and TOOT objects which are not listed in NED,
  see text for details. The errors associated with the flux densities are
  typically a few per cent.}
\label{tab:RGs}
\begin{tabular}{lrrr}
\hline
     Name & S$_{325}$ (Jy) & $\alpha$ & z \\    
\hline
              3C\ 175.1  &  6.939  &  0.85  &  0.920 \\
               3C\ 184    &  9.097  &  0.87  & 0.994 \\
               3C\ 22   &  8.348   &  0.90   &  0.936 \\
               3C\ 268.1 & 15.615  &  0.58   &  0.970 \\
               3C\ 280  &  16.025 &  0.81  &  0.996 \\
               3C\ 289  &  8.278  &  0.84  &  0.967 \\
               3C\ 343  &  13.413 &  0.68  &  0.988 \\
               3C\ 356  &  6.820  &  1.04  &  1.079 \\   
          6C\ E0943+3958  &  1.182  &  0.85  &  1.035  \\
          6C\ E1011+3632  &  1.190  &  0.79  &  1.042  \\
          6C\ E1017+3712  &  1.540  &  1.00  &  1.053  \\
          6C\ E1019+3924  &  1.690  &  0.94  &  0.923 \\
          6C\ E1129+3710  &  1.543  &  0.89  &  1.060  \\
          6C\ E1212+3805  &  1.408  &  1.06  &  0.95   \\
          6C\ E1217+3645  &  1.402  &  0.94  &  1.088  \\
          6C\ E1256+3648  &  1.760  &  0.81  &  1.07   \\
          6C\ E1257+3633  &  1.036  &  1.08  &  1.004  \\
          6C*0128+394     &  1.322  &  0.50  &  0.929  \\
          6C*0133+486     &  0.742  &  1.22  &  1.029  \\
          5C\ 6.24  &  0.839  &  0.77  &  1.073 \\
          5C\ 7.17  &  0.469  &  0.93  &  0.936 \\
          5C\ 7.23  &  0.546  &  0.78  &  1.098 \\
          5C\ 7.242  &  0.304  &  0.94  &  0.992 \\
          5C\ 7.82  &  0.371  &  0.93  &  0.918 \\
          TOOT1066  & 0.098  & 0.87  & 0.926\\
          TOOT1140  & 0.298  & 0.75 & 0.911\\
          TOOT1267  & 0.282  & 0.80 & 0.968\\

\hline
\end{tabular}
\end{table}

\begin{figure}
\centering
\includegraphics[width=0.9\columnwidth]{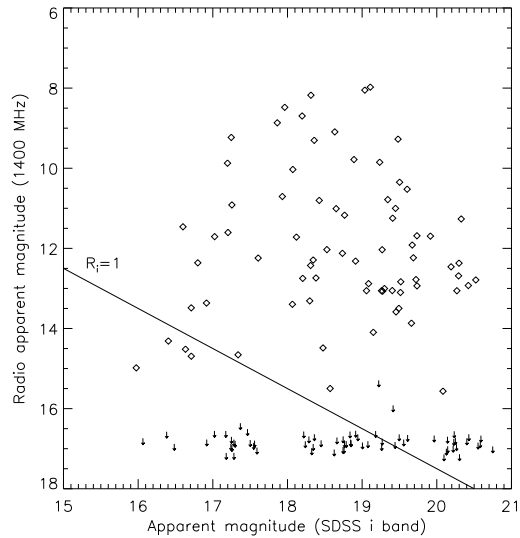}
\caption{Optical apparent magnitude (SDSS i band) vs radio apparent magnitude
  (NVSS 1400~MHz) for the quasar samples. RLQs are plotted as diamonds while
  RQQs are shown as upper limits. The line shows the parameter $R_{\rm i} = 1$
  \citep{Ivezic} which is used to determine radio-loudness, i.e. objects
  falling above the line are classified as radio-loud while objects falling
  below the line are classified as radio-quiet. The plot shows that, by this
  definition, all but 4 of our RLQs would be classified as radio-loud and that
  {\em at least\/} two thirds of our RQQs would be classified as radio-quiet.}
\label{fig:radioloudness}
\end{figure}

\begin{figure}
\centering
\includegraphics[width=0.7\columnwidth]{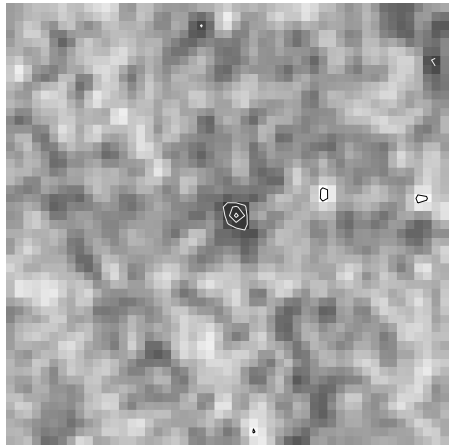}
\caption{Stacked and averaged radio image of the positions of our RQQs using
  data from the FIRST survey. The contours show the 3-, 4- and 5-$\sigma$
  levels. Black contours are negative and white positive. The image dimensions
  are $1.5\times1.5$ arcmin.}
\label{fig:RQQimage}
\end{figure}

\begin{figure}
\centering
\includegraphics[width=0.9\columnwidth]{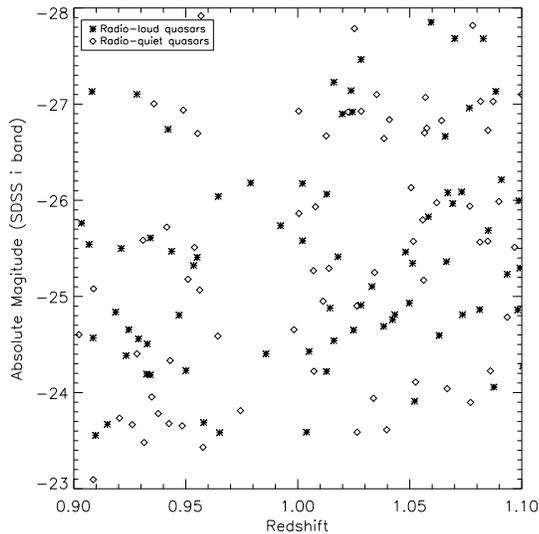}
\caption{Optical absolute magnitude (SDSS i-band) vs redshift for
  the quasars. RLQs are shown with asterisks and RQQs with diamonds.}
\label{fig:Mag_z}
\end{figure}

\begin{figure}
\centering
\includegraphics[width=0.9\columnwidth]{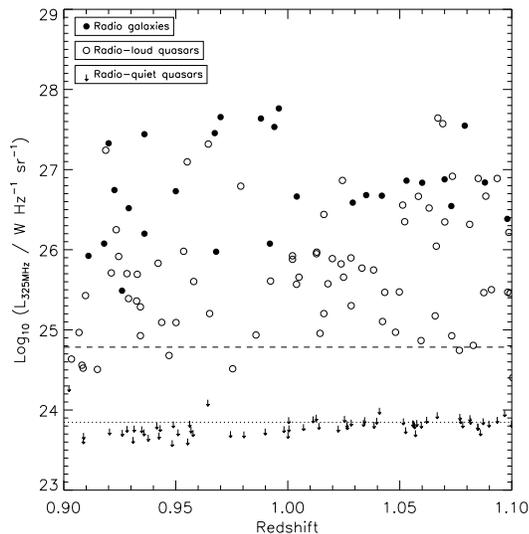}
\caption{Low frequency radio luminosity vs redshift for our
  sample. RLQs are shown with open circles and RGs with filled circles
  (data are rest-frame 325~MHz from WENSS). For the RQQs, 5-$\sigma$
  upper limits (extrapolated to rest-frame 325~MHz) from the FIRST
  survey are shown. Where WENSS data were unavailable for the RLQs due
  to sky coverage (about 10 objects) the 325~MHz flux density was
  extrapolated from the NVSS survey at 1400~MHz assuming
  $\alpha=0.7$. The dashed line shows the average 5-$\sigma$ limit of
  the WENSS survey, converted to a luminosity at $z=1$ by assuming
  $\alpha=0.7$; the RLQs were selected to have radio luminosities
  falling above this line. The dotted line shows the average
  5-$\sigma$ limit of the FIRST survey, extrapolated to 325~MHz and
  again converted to a luminosity; the RQQs were selected to have a
  radio luminosity falling below this line. The assumed spectral
  indices for some conversions explains why some objects fall between
  the lines on this plot.}
\label{fig:radioL_z}
\end{figure}

\subsection{Selection}

\label{section:sample}

The quasars were selected by their optical colours in the SDSS Quasar Survey
\citep{Schneider05}. Using the SDSS to select the quasars allowed us to select
a large enough initial sample that the RLQ and RQQ samples were chosen in
identical ways. The initial sample that met the SDSS colour criteria for
quasars was then cross referenced with the NRAO VLA Sky Survey (NVSS;
\citealt{Condon98}), the VLA FIRST survey (\citealt{Becker95}) and the
Westerbork Northern Sky Survey (WENSS; \citealt{Rengelink97}) to pick out the
RLQs and RQQs.

RLQs were chosen to have a low frequency WENSS (325~MHz) flux density of
greater than 18~mJy which is the 5-$\sigma$ limit of the survey. At $z\sim1$
this corresponds to a radio luminosity almost entirely within the radio-loud
domain. Fig.~\ref{fig:radioloudness} compares our sample to an alternative
definition of radio-loudness used by \cite{Ivezic}. Here radio-loud objects are
defined to have $R_{\rm i}>1$ where $R_{\rm i}={\rm log}(F_{\rm radio}/F_{\rm
  i})$ and $F_{\rm radio}$ and $F_{\rm i}$ are flux densities measured at
1.4~GHz and in the $i$-band respectively (K-corrections are not applied). With
the exception of 4 objects, all of our RLQs would be considered radio-loud
using this definition. We note that these four objects have only one flux
density measurement at radio wavelengths, and hence may in fact fall above the
line if their spectral indices differ from the value we have assigned to them,
i.e. $0.7$ which is the mean value of the measured spectral indices. Using a
low frequency radio flux to define the RLQs allows them to be compared more
easily to the RGs without a severe orientation bias.

The RQQs were defined as being undetected by the FIRST survey at the 5-$\sigma$
level. FIRST was used for this definition because it provides a more sensitive
flux density limit than WENSS. In this case, the RQQs are not selected to be
true radio-quiet objects as defined by the radio-loudness parameter for
example. In order to get an estimate for the average radio power of the RQQs in
our sample we have used a technique of stacking the radio images of
non-detections to reveal the average value of the FIRST radio power
(e.g. \citealt{White07}). In essence it involves stacking up the radio images
at the known positions of the RQQs, weighting each image by its standard
deviation and then computing the average radio emission or obtaining a
sensitive upper limit. The stacked image is shown in
Fig.~\ref{fig:RQQimage}. Using this technique we find an average flux density
for our RQQs at 1.4~GHz of $0.10\pm0.02$~mJy (i.e. a 5-$\sigma$ detection).
Assuming a spectral index of 0.7 allows us to extrapolate to a 325~MHz flux
density of $0.30\pm0.06$~mJy which at $z\sim1$ corresponds to a 325~MHz
luminosity, Log$_{10}(\rm L_{325}/ \rm{W\ Hz}^{-1}\ \rm{sr}^{-1})=23.02$.

Lists of around 75 RLQs and 71 RQQs were chosen for observation to be matched
in optical luminosity and span the full 5 optical magnitudes available; the
selected sources are shown in bold in Fig.~\ref{fig:SDSS}. The distribution of
optical magnitudes within the selected redshift range is shown in
Fig.~\ref{fig:Mag_z}.  The RGs were selected from the low frequency, (178 or
151~MHz; orientation independent) radio samples 3CRR \citep{LaingLongair83},
6CE \citep{Eales85}, 7CRS \citep{Willott98} and TOOT surveys
\citep{Hill&Rawlings03}. Together, these surveys give 27 RGs in the same $0.9 <
z < 1.1$ redshift range as our quasars. The reason therefore for our
substantially smaller RG sample is purely due to the limit of the known RG
population at $z\sim1$.

The RLQ and RG radio luminosity distribution within the selected redshift range
is shown in Fig.~\ref{fig:radioL_z} which shows that, on average, the RGs are
more radio-luminous than the RLQs, albeit with a significant overlap.  Using
the FIRST radio images we can also place an upper limit on the radio emission
of each RQQ; see Fig.~\ref{fig:radioloudness} and Fig.~\ref{fig:radioL_z},
which shows that at least two thirds of our RQQs and most likely more would be
classified as radio-quiet using the definition from \cite{Ivezic}. In
comparison to these limits the least radio-loud RLQ has a 325~MHz luminosity of
Log$_{10}(\rm L_{325}/ \rm{W\ Hz}^{-1}\ \rm{sr}^{-1})=24.5$, showing that there
is at least an order of magnitude difference (several between the means) in the
radio emission of our RQQs and RLQs.  The reason for the gap between the RLQs
and RQQs in radio luminosity seen in Fig.~\ref{fig:radioL_z} is due to the
difference in the survey depths of the WENSS and FIRST surveys from which they
were selected. The 5-$\sigma$ limit used for the RLQ lower limit and the RQQs
upper limit are not identical, as shown by the dashed and dotted lines in
Fig.~\ref{fig:radioL_z}. This leaves a small region on the radio luminosity
axis uncovered by our data, but it is important to note this is due to our
selection rather than evidence for a real radio power dichotomy.

\subsection{Observations and data reduction}

We omitted from our {\em Spitzer\/} observations all targets which lie
within the overlap regions of the SDSS, the Spitzer Wide-Area Infrared
Extragalactic Legacy Survey (SWIRE; \citealt{Lonsdale03}) and the
Extragalactic First Look Survey (XFLS; \citealt{Lacy05}) regions. This
was because \cite{Richards06} have already compiled the data for these
objects, of which two are RLQs and 15 are RQQs. We also found that two
of the 3C RGs have adequate data in the archive: 3C\ 356 (ID3329; PI
Stern) and 3C\ 184 (ID17; PI Fazio). These data were downloaded
  from the {\em Spitzer\/} archive and added to our own observations
  which are described below.

Our IRAC observation were carried out in all four bands between 2006 August and
2007 August. We performed 5pt Gaussian dithers with the medium cycling pattern
and 12~sec frame-time to ensure good scattered light rejection and good
photometry. The data were reduced with the standard pipeline version S15.0.5
giving final maps with a pixel scale of 1.2 arcsec. In this paper we use the
$3.6~\mu m$ channel to study the environments of our AGN since it samples light
emitted at $1.8~\mu m$ at $z\sim1$ which is close to the peak of the stellar
emission from galaxies. It is also the most sensitive IRAC channel.  An
aperture correction of 1.48 was applied to the $3.6~\mu m$ flux densities, as
determined by the {\em Spitzer\/} First Look Survey \citep{Lacy05}. Full
details of the observations and data reduction are given in Jarvis et al. , in
prep.

\section{Source Extraction}

The images were cut down using {\sc iraf} tasks to leave $3.4 \times 3.4$
arcmin square images centered on the position of the AGN. This process removes
any edge effects and underexposed edges caused by the dither pattern, but
leaves enough area to sample the AGN environment out to a radius of 800~kpc at
$z\sim1$. To create catalogues of the images we used the {\sc SExtractor}
software package \citep{Bertin&Arnouts96}. We used a detection threshold of 3
adjacent pixels each at $1.5~\sigma$ above the RMS background level. The {\sc
  seeing$\_$fwhm} was set to 1.67~arcsec which the measured resolution of our
images. On inspection of the output catalogues it was found that the default
value for the deblending parameter (0.005) resulted in non-detection of faint
sources close to bright AGN and stars. We therefore adopted the value used by
\cite{Lacy05} of 0.0001 which gave a significant improvement, although in a
small number of cases it led to the inclusion of spurious sources identified
with diffraction spikes.

To check for spurious sources we inverted the maps and ran the same source
extraction configuration. The results showed that 96 per cent of the fields had
no spurious 5 sigma sources. However, the remaining 4 per cent had spurious
sources associated with the diffraction spikes of bright stars (which give
pronounced negative artifacts). We therefore used this method to indentify
those images affected by diffaction spikes and then manually checked their
non-inverted catalogues for sources that were obviously spurious, these were
then removed from the catalogues.

\subsection{Source cuts}

\label{section:cuts}

The catalogues were filtered to reduce the noise produced by spurious and
foreground sources. Filtering criteria were applied identically to the AGN
fields and the blank fields (see below).

\begin{figure}
\centering \includegraphics[width=0.9\columnwidth]{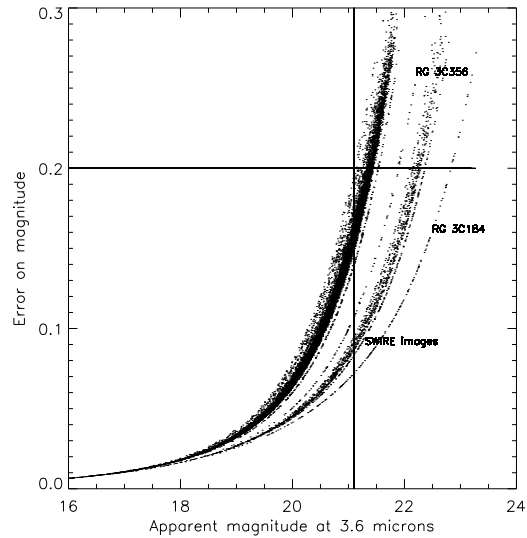}
\caption{The apparent magnitudes of all sources at 3.6~$\mu$m vs the error on
  that magnitude. The deeper SWIRE images are labelled, and the lines show the
  0.2 magnitude error ($5~\sigma$) level and the corresponding magnitude for
  the shallowest image, i.e. $21.1$ mag. The two other smaller
  bands shown and labelled are from the two RGs which had already been imaged by
  other programmes; see the text for details.}
\label{fig:5sigmalevel}
\end{figure}

In order to ensure that all images were of the same sensitivity a cut was made
of source detections below a conservative 5-$\sigma$ level (see
Fig.~\ref{fig:5sigmalevel}). This level is defined as the 5-$\sigma$ level of
the shallowest image in the sample thus ensuring that any sources used could be
detected in any image. The cut is especially necessary because of the use of
images from SWIRE and the archive which are deeper than our own. Using this
method, all sources in our fields with apparent magnitudes fainter than 21.1
(13.1 $\mu$Jy) were excluded from our analysis. To put this limit into context,
the absolute magnitude of the break in the $K$-band galaxy luminosity function
at $z\sim1$ is $K^* = -23.0$ (\citealt{Cir09}). An elliptical galaxy at
$z\sim1$ has $K-3.6=0.27$ (\citealt{BC03}) giving an equivalent break in the
3.6~$\mu$m luminosity function of $-23.3$ or an apparent magnitude of
$m_{3.6}^*\sim20.8$. Thus at the 5-$\sigma$ depth of our survey we are
sensitive to galaxies with 3.6~$\mu$m luminosities of $\sim L^*$ or greater. We
are sampling star-light emitted longwards of the Balmer break at $z\sim1$ so
any companions to the AGN are likely to be galaxies with a substantial old
stellar population.

In order to remove stars the {\sc SExtractor class$\_$star} output parameter
was used. This returns a value between 1 and 0 where 1 is a perfectly star-like
object and 0 is very non-star-like. We have cut all sources that have a {\sc
  class$\_$star} parameter greater than 0.8, as used by \cite{Best03} and
\cite{Smith00}. We also ran the same analysis using 0.95, another commonly
adopted value, and found almost identical results, suggesting that most objects
cut at the 0.8 level had values greater than 0.95.

\section{Analysis}

Once the catalogues were created they were searched for sources in annuli
working out from the target AGN. The annuli were kept to a fixed area, rather
than fixed width; this keeps the signal-to-noise and Poisson errors of similar
size from bin to bin and also allows for a larger number of annuli. The target
AGN is excluded from the galaxy counts as this would bias the results towards
there being an over-density in the first bin.

To get an estimate for the average number of counts in the field to compare
with the AGN fields we have made use of the extra adjacent field that comes
with {\em Spitzer\/} images. This is a result of the way {\em Spitzer\/}
works. IRAC has four detectors but can only point two at the target at any one
time so the telescope has to offset to allow the other two detectors to image
the target. While this happens the untargeted detectors image a region of an
adjacent field to the same depth as the target fields. As these fields are not
targeted at the AGN they can be used as blank or control fields, and were thus
treated in exactly the same way as the AGN fields. Hence the region with the
same exposure time as the main fields was source extracted in the same manner
as the AGN fields and the average source density was computed.

We can also use the blank fields to get a measure of the local foreground and
background in each region and subtract the mean source density in each blank
field from its AGN field. In this case, we are measuring the source
over-density rather than just the source density, which better allows us to
stack the results. One possible pitfall of this approach is that the blank
fields might be close enough to the AGN that any over-density will extend into
them. The proximity of blank fields to the target fields is therefore a
trade-off between the desire to subtract a local foreground whilst not wanting
to be so close that any over-density is also subtracted.

For comparison we have calculated the global source density level for the SWIRE
fields using the same source cuts described above.  Our background source
density is $7.34\pm0.35$ arcmin$^{-2}$ which compares to the overall SWIRE
average from the 3 northern fields (covering $\sim25$~deg$^2$) of $6.79\pm0.34$
arcmin$^{-2}$. The quoted uncertainties in these backgrounds are calculated by
adding the Poisson and cosmic variance errors in quadrature. Cosmic variance
errors were calculated by placing 173 tiles on to the SWIRE fields, each having
an equal area to one of our blank fields to give a mock survey of equivalent
area to that of our real survey. We repeated this procedure 43 times and
calculated the mean background level and its standard deviation which is the
cosmic variance error.  At face value, our fields have a slightly higher
average background level than the SWIRE fields but they are entirely consistent
within the errors.

Does this analysis imply that any over-density surrounding our AGN extends into
the blank fields? This is unlikely for the following reasons. Firstly, there is
evidence in the literature that over-densities for the most powerful radio
galaxies (at $z\sim1.6$) extend out to, at most, 1.6 Mpc (\citealt{Best03})
whereas our blank fields are at their closest point $\sim$2.8~Mpc (at z$\sim$1)
from the AGN. Secondly, when we break the blank fields into strips we see no
increase in source density towards the AGN. It is possible that the increased
background could be due to stars that are not removed by the {\sc class\_star}
parameter cut. This may be a larger effect in our images since the SWIRE fields
are confined to high Galactic latitudes which are often pre-selected to have
low foregrounds whereas our fields are more evenly spread over the northern
SDSS regions.

\subsection{Completeness}

\label{section:completeness}

In order to measure and correct for the completeness of our data we conducted
extensive simulations to add and recover artifical sources. We added 1000
sources in to each of our AGN fields for 30 flux bins (i.e. 30000 sources per
image) and proportionally less for the blank fields as they are of a smaller
area. These sources were chosen to be Gaussian with a FWHM of 2 arcsec, and
were scaled to the required flux. They were added across each image in a grid
pattern in batches of 100 for the AGN fields and proportionally less for the
blank fields. The sources were allowed to randomly move around within the grid
as far as possible without two sources ever being placed close enough as to not
be deblended. Sources were considered to be recovered if they were found in the
{\sc SExtractor} catalogues within 1.5 pixels (1.8 arcsec) of their input
position and had an extracted flux within a factor of 2 of the input value. In
the AGN fields we computed the completeness for each annulus individually which
allows us to correct for bright objects masking regions of certain annuli. In
the blank fields we used the average completeness for the whole blank field
region as we are only interested in attaining one value per field.

\begin{figure}
\centering
\includegraphics[width=0.9\columnwidth]{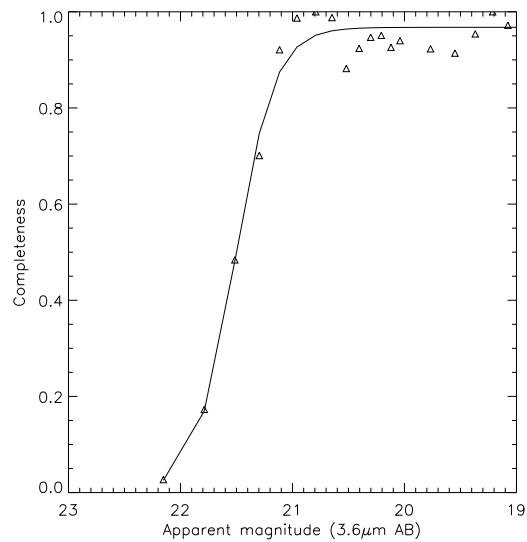}
\caption{An example of a completeness curve with empirical model fit. Our flux
  limit is 13~$\mu$Jy which is therefore the lowest flux at which we use a
  completeness correction.}
\label{fig:example_fit}
\end{figure}

To eliminate the scatter in the measured completeness curves we then fitted
them with an empirical model of the form $Completeness=(S^a)/(b+cS^a)$
(\citealt{Cop06}) where $S$ is the 3.6-$\mu$m flux density and $a$, $b$ and $c$
are constants that are fitted. An example of one of these fits with the
overlaid data points is shown in Fig.~\ref{fig:example_fit}. This process then
allowed us to apply a different completeness correction to every annulus in
each image as a function of flux which corrects for any lost area due to bright
stars or diffraction spikes as well as sources lost due to being in crowded
regions.

\begin{figure}
\centering
\includegraphics[width=0.9\columnwidth]{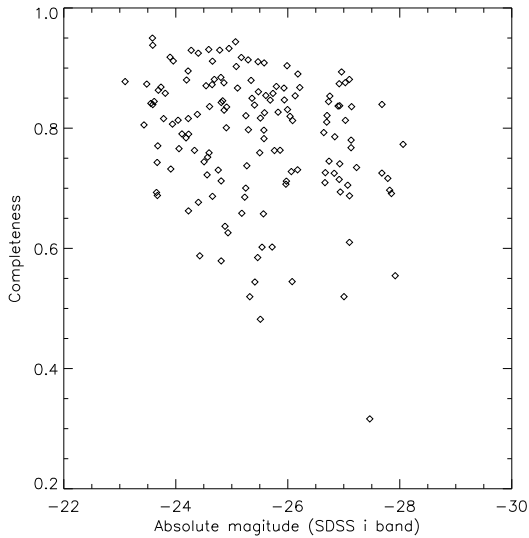}
\caption{Plot showing the optical absoulute magnitude (SDSS i band) vs the
  completeness at our flux limit (13~$\mu$Jy) in the first annulus for all the
  quasars in our sample. Correlation analysis shows a
  correlation at the 99.7~per cent level.}
\label{fig:completeness}
\end{figure}

As expected, the completeness is lower than average in the first annulus of the
quasar fields which is due to the bright AGN hindering the detection of faint
sources (see Fig.~\ref{fig:completeness}). This effect was also noted by
\cite{Yee84}. Indeed, we find a highly significant correlation between the
optical magnitude of the AGN and the completeness of the first bin for the
quasars at the 99.7~per cent level using correlation analysis (Spearman rank
and Kendall tau). Hence applying the completeness corrections was found to
boost the source density in the first annulus relative to the others (although
the significance is not changed as the error is also scaled by the
completeness). This effect is not as prominent in the RG fields as they are
generally less luminous in the optical/infrared, due to the quasar nucleus
being obscured so that we only see the stellar light from the host galaxy.

\begin{figure}
\centering
\includegraphics[width=0.8\columnwidth]{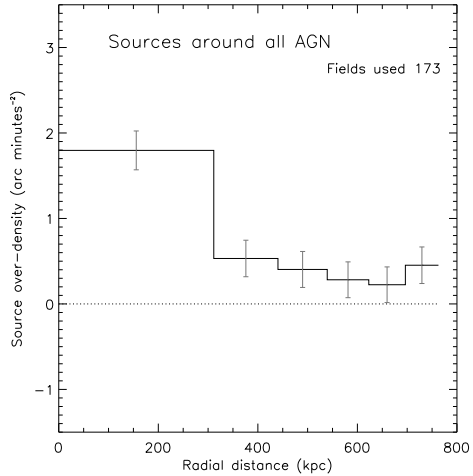}
\caption{Histogram showing the average source over-density in all the AGN
  fields after they were corrected for completeness and the local background
  has been subtracted. The error bars for each field are the Poisson error for
  each bin combined in quadrature with the Poisson error on the blank field
  level that was subtracted, both scaled by their mean completeness correction;
  these values are then added in quadrature to give the final error bar on the
  stacked histogram. The dotted line simply highlights the zero level.}
\label{fig:all_AGN}
\end{figure}

\section{Results}

\begin{table}
\centering 
\caption{The over-density shown in Fig.~\ref{fig:all_AGN} tabulated for each
  bin and as an average for the outer bins (i.e. excluding bin one). The table
  shows the over-density ($\Delta N$), the over-density error (err $\Delta N$)
  and the Sigma value which is the number of 1-sigma error bars (err $\Delta
  N$) that the over-density ($\Delta N$) is above zero.}
\label{tab:all_AGN}
\begin{tabular}{cccc}
\hline
Bin & $\Delta N$ & err
$\Delta N$ & Sigma \\  & (arcmin$^{-2}$)
&(arcmin$^{-2}$) & \\
\hline
1 &  1.80  &  0.21  &   8.50  \\
2 &  0.53  &  0.20  &   2.71  \\
3 &  0.40  &  0.19  &   2.08  \\
4 &  0.28  &  0.19  &   1.46  \\
5 &  0.22  &  0.19  &   1.17  \\
6 &  0.45  &  0.20  &   2.29  \\
Outer bins &  0.38  &  0.09  &   4.26  \\
\hline
\end{tabular}
\end{table}

Conducting the radial search on individual fields does not give significant
results because the IRAC data are dominated by foreground sources. There are
also large Poisson errors resulting from counting small numbers of
objects. However, these problems are reduced if multiple fields are stacked up
because the larger number of counts involved reduces the associated Poisson
errors, as well as averaging out the variations in foreground sources. Hence in
this paper we concentrate mainly on a statistical analysis of stacked data from
multiple fields.

\begin{figure*}
\centering 
\includegraphics[width=0.3\linewidth]{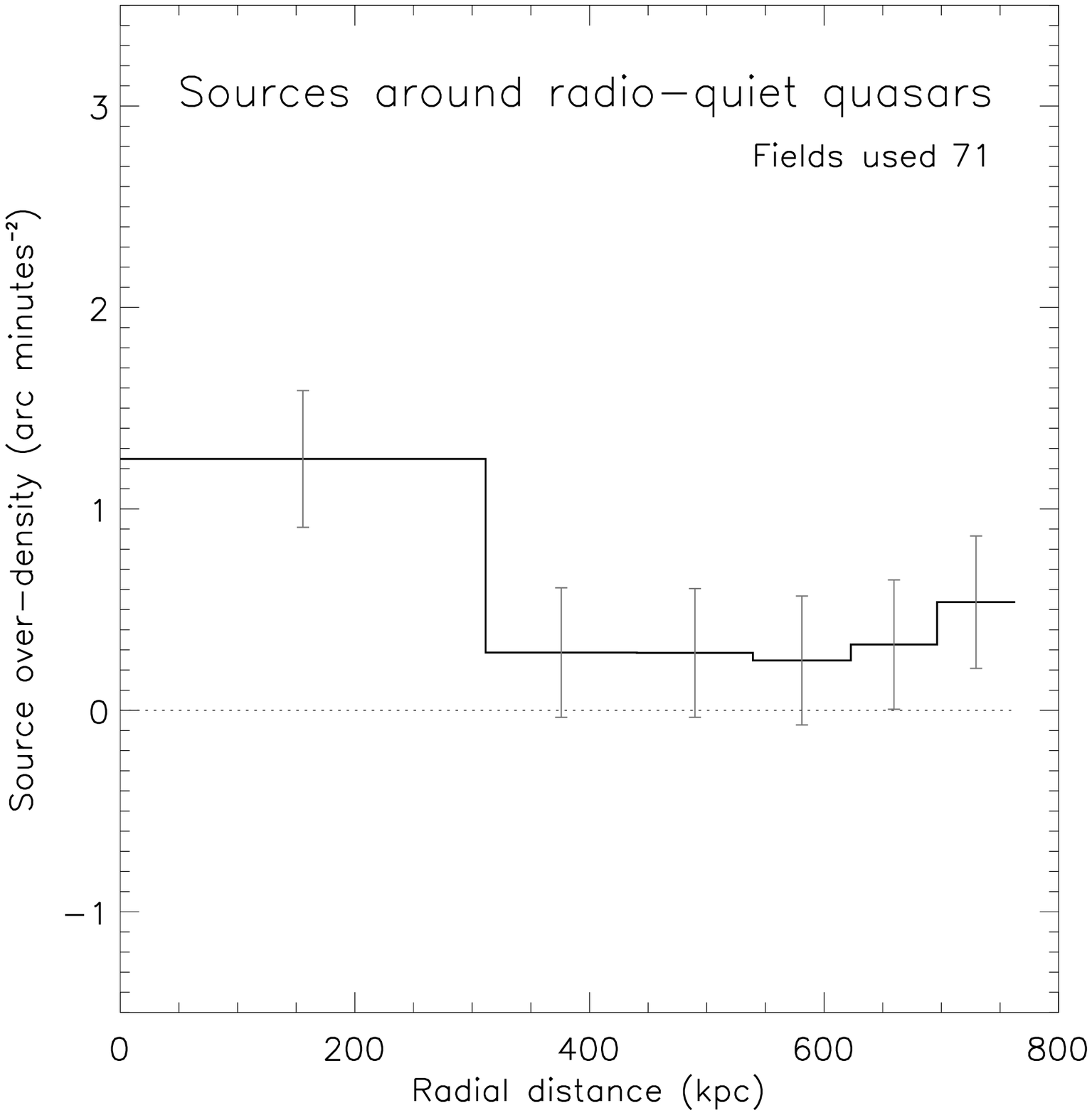}
\includegraphics[width=0.3\linewidth]{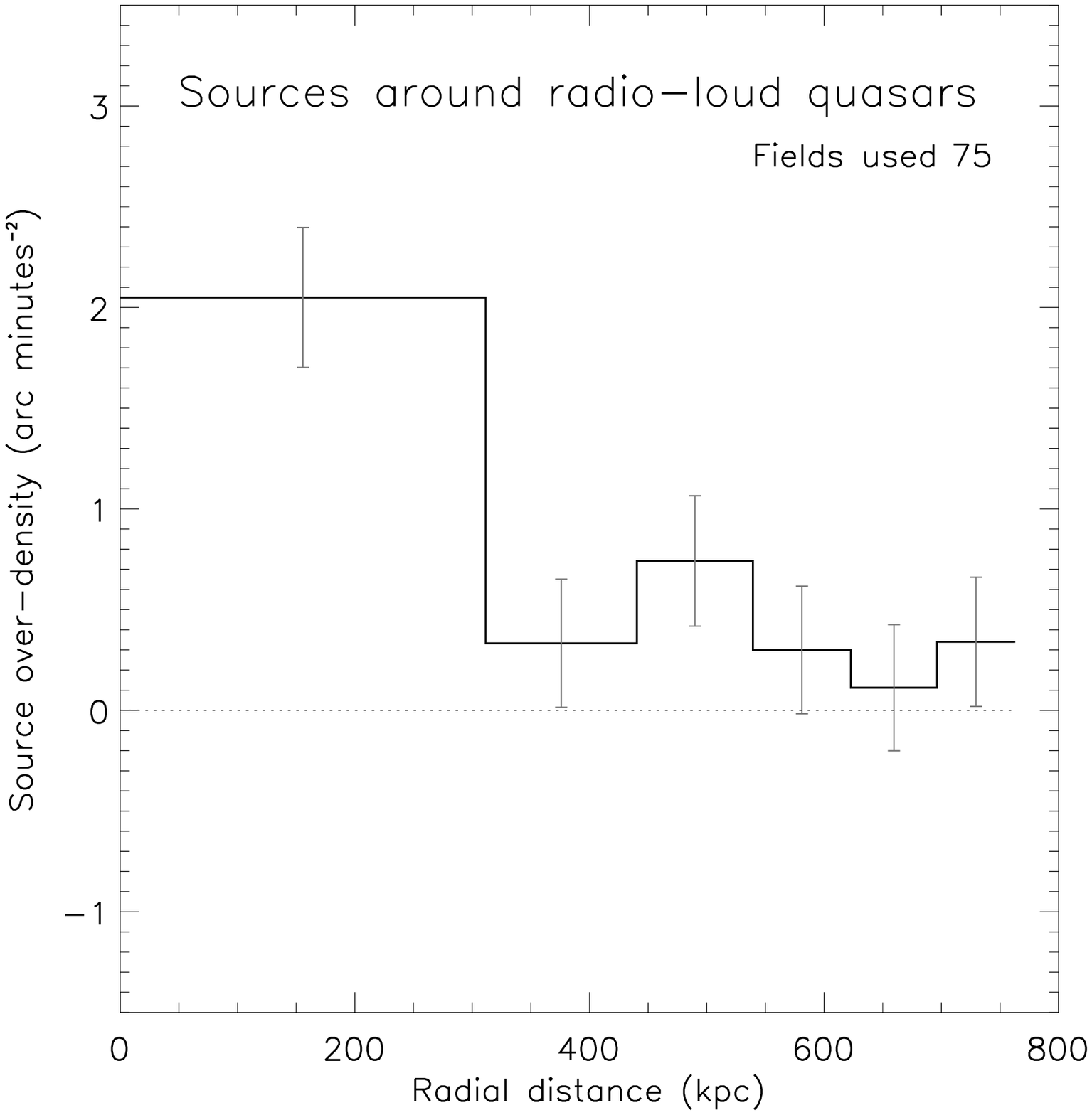}
\includegraphics[width=0.3\linewidth]{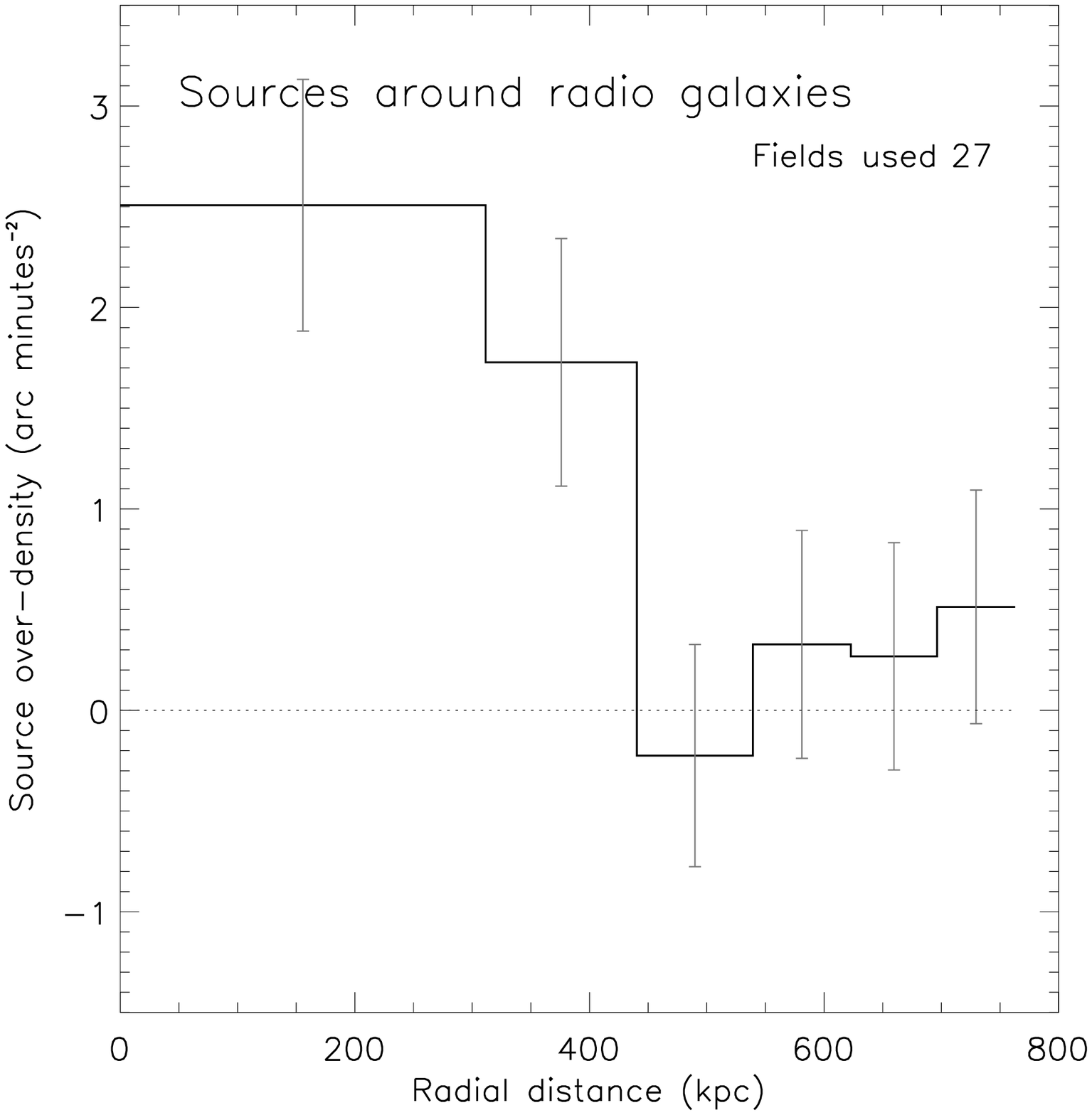}
\caption{Histograms showing the averaged source over-density for the AGN
  sub-samples. The right-hand panel shows the RQQs, the middle panel shows the
  RLQs and the left-hand panel shows the RGs. The error bars for each field are
  the Poission errors for each bin combined in quadrature with the Poisson
  error on the blank field level, that was subtracted, both scaled by their
  mean completeness correction; these values are then added in quadrature to
  give the final error bar on the stacked histogram. The dotted lines simply
  show the zero levels.}
\label{fig:types}
\end{figure*}

\subsection{Over-density for the whole AGN sample}

\label{section:all}

In Fig.~\ref{fig:all_AGN} we show the results of a radial search conducted on
all our AGN fields, centered on the AGN, stacked together and averaged. Note
that the local background level calculated from the blank field of each AGN has
been subtracted so the plot shows source over-density rather than source
density. The error bars for each field are the combined Poisson errors of the
count in the AGN field added in quadrature to the Poisson error of the blank
field value that was subtracted, both scaled by their mean completeness
correction; these are then added in quadrature to give the final error bar on
the stacked histogram. It is obvious that there is an over-density in the AGN
fields as a whole, since all of the bins have a nominal value above the
background level, but also a big increase in this excess towards the position
of the AGN.

It thus appears that the bulk of the over-density is concentrated within the
first bin or within 300~kpc (physical units) of the target AGNs. The chosen bin
size came from a trade off between gaining a reasonable signal-to-noise in the
first bin whilst not watering down the over-density by including too large an
area. The radial search was thus repeated using annuli of varying size and it
was found that the optimum size is that adopted, which has an area of
1.4~arcmin$^{2}$. This exercise also showed that, even when smaller bins were
used, the central over-density still extended to $200-300$~kpc before dropping
towards the background level.

To calculate the significance of the over-density in each bin, we simply use
the number of 1-$\sigma$ error bars that the over-density is above zero. This
is shown in Table~\ref{tab:all_AGN} which shows the relevant values for each
annular bin of the histogram shown in Fig.~\ref{fig:all_AGN}. It can be seen
that if we just consider the first bin this is over-dense at the 8.5-$\sigma$
level.

In order to test whether the over-density in the outer bins (excluding bin one)
is significant we added the counts in the outer bins and calculated the
combined Poisson error. This gives an over-density of
$0.37\pm0.09$~arcmin$^{-2}$ and a significance of 4.3~$\sigma$ (see
Table~\ref{tab:all_AGN}). Therefore, the outer bins are also significantly
over-dense (out to at least 800~kpc) but the bulk of the over-density is mainly
found in the first bin. This pattern of a sharp peak in the central source
over-density and then an extended flatter over-density was also reported by
\cite{Best03} for powerful radio galaxies at $z\sim1.6$, and by \cite{Serber06}
for $M_{\rm i} \leq -22$, $z\leq0.4$ SDSS quasars.

\begin{table}
\centering 
\caption{The over-densities from Fig.~\ref{fig:types} shown for the first three
  bins and as an average for the outer bins (excluding bin 1). The table shows
  the over-density ($\Delta N$) and the sigma value which is the number of
  $1~\sigma$ errors that the source density is above the blank field source
  density.}
\label{tab:types}
\begin{tabular}{cccc}
\hline
AGN type & Bin & $\Delta N$ & Sigma \\ && (arcmin$^{-2}$) & \\
\hline
Radio & 1 & 2.51 & 4.01 \\
galaxies & 2 & 1.73 & 2.81 \\
& 3 & -0.22 & -0.41 \\
& Outer bins & 0.52 & 2.15 \\
Radio-loud & 1 & 2.05 & 5.90 \\
quasars & 2 & 0.33 & 1.05 \\
& 3 & 0.74 & 2.29 \\
& Outer bins & 0.37  & 2.73 \\
Radio-quiet & 1 & 1.25 & 3.67 \\
quasars & 2 & 0.29 & 0.89 \\
& 3 & 0.28 & 0.89 \\
& Outer bins & 0.34  & 2.48 \\
\hline
\end{tabular}
\end{table}

\subsection{Over-density versus AGN Type}

To investigate trends with AGN classification we have divided the AGN into RQQ,
RLQ and RG sub-types and performed a similar analysis to that described in
Section \ref{section:all}. The results of this analysis are shown in
Fig.~\ref{fig:types}. On first inspection, the most notable difference between
the subsets is the apparently larger over-density in the RLQ and RG samples
when compared to the RQQ sample.

Using the same method as for the whole sample described in Section
\ref{section:all}, we have calculated the significance of the over-density in
each annulus for each of these sub-samples. The main results are summarized in
Table~\ref{tab:types} which shows the over-densities in the first three bins
and their significances and the same values in just the outer bins (excluding
bin one). Focusing on the first bin, which in all cases appears to contain the
bulk of the over-density, we find over-densities for the RGs of 4.0~$\sigma$,
for the RLQs of 5.9~$\sigma$ and for the RQQs of 3.7~$\sigma$. The lower
significance level for the RG sample probably just reflects the small sample size.
It is worth noting that in this
subset the second bin is also significant at the 2.8~$\sigma$ level, which if
combined with the first bin would give a more significant over-density.

To quantify the difference we see in the first bins of the RLQs and RQQs we
conducted a Mann-Whitney test, which is a non-parametric form of a $t$-test,
and gives the probability that two data sets, in our case the source
over-densities around our AGN, have the same mean. Considering source densities
in the first bin this test indicates that the RLQs inhabit, on average, more
dense environments than the RQQs at the $>96$~per cent confidence level. It
therefore appears that all of our AGN have a significant over-density out to
$\sim$300~kpc with the radio-loud objects, on average, having larger
over-densities. Interestingly the RGs appear to have a larger central
over-density than in the quasar fields although more RG data are needed to
confirm this tentative result.

In order to test that these results are not being caused by the presence of a
few very over-dense fields which happen to fall into our radio-loud samples we
removed the 5 most over-dense fields from our RLQs and the five least
over-dense fields from our RQQs. The observed trend for the first bin still
holds, although obviously there is a drop in significance due to removal of the
most extreme fields in each case. It is clear from this analysis, however, that
there is a fair amount of field-to-field variation which could be attributed to
either actual changes in the enviromental richness between fields or to
field-to-field changes in the foreground/background contamination.

\subsection{Black-hole mass vs environmental density}

In order to try to understand the contrast between the fields of the radio-loud
and radio-quiet objects we can use estimates of black-hole mass to search for
possible trends. For example, are the biggest over-densities found around the
AGN which host the largest black holes?

Our black-hole mass estimates are computed using the virial estimator
and the \mg line at $2800$\,\AA  using SDSS spectroscopy, a
technique described by \cite{McLure&Jarvis02}, and based on work by
\cite{McLure04}. Note that we do not have black-hole mass estimates
for the radio galaxies because the broad-line region is obscured in
these objects, but estimates are available for all of the quasars.\\

\begin{figure}
\centering
\includegraphics[width=0.8\columnwidth]{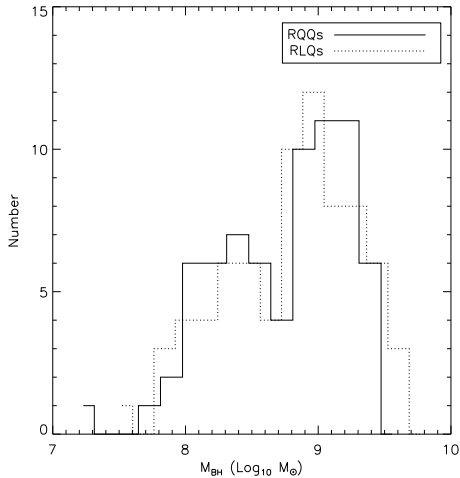}
\caption{Histograms showing the distributions of RLQ and RQQ
  black-hole mass estimates.}
\label{fig:Mbh-hist}
\end{figure}

To test whether the difference we find between the fields of the RLQs and RQQs
might be related to intrinsically different black-hole mass distributions, as
suggested in previous papers (e.g. \citealt{Lacy01};
\citealt{McLure&Jarvis04}), we again conducted a Mann-Whitney test. The test
suggested that the mean black-hole masses for the two samples are not
significantly different; the returned probability that they have the same mean
is 0.32. Moreover, a two-sample Kolmogorov-Smirnov (K-S) test returns a
probability of 0.90 that the samples are drawn from the same parent
distribution.  We stress that this result does not contradict previous work but
is rather a direct consequence of the manner in which our samples were
initially matched in absolute optical magnitude and colours as well as the
relatively small sample size.  The mean black-hole masses are $\langle{\rm
  Log}_{10}(M_{\rm BH}/\Msolar)\rangle=8.87\pm0.06$ for the RLQs and
$\langle{\rm Log}_{10}(M_{\rm BH}/\Msolar)\rangle=8.81\pm0.06$ for the RQQ so
the means of the two samples are consistent and well within 1-$\sigma$ of each
other. In any case the real uncertainties are likely to be bigger once
systematics are taken into account.  The distribution of black-hole masses is
shown for both samples in Fig.~\ref{fig:Mbh-hist}.\\

\begin{figure*}
\centering

\includegraphics[width=0.8\columnwidth]{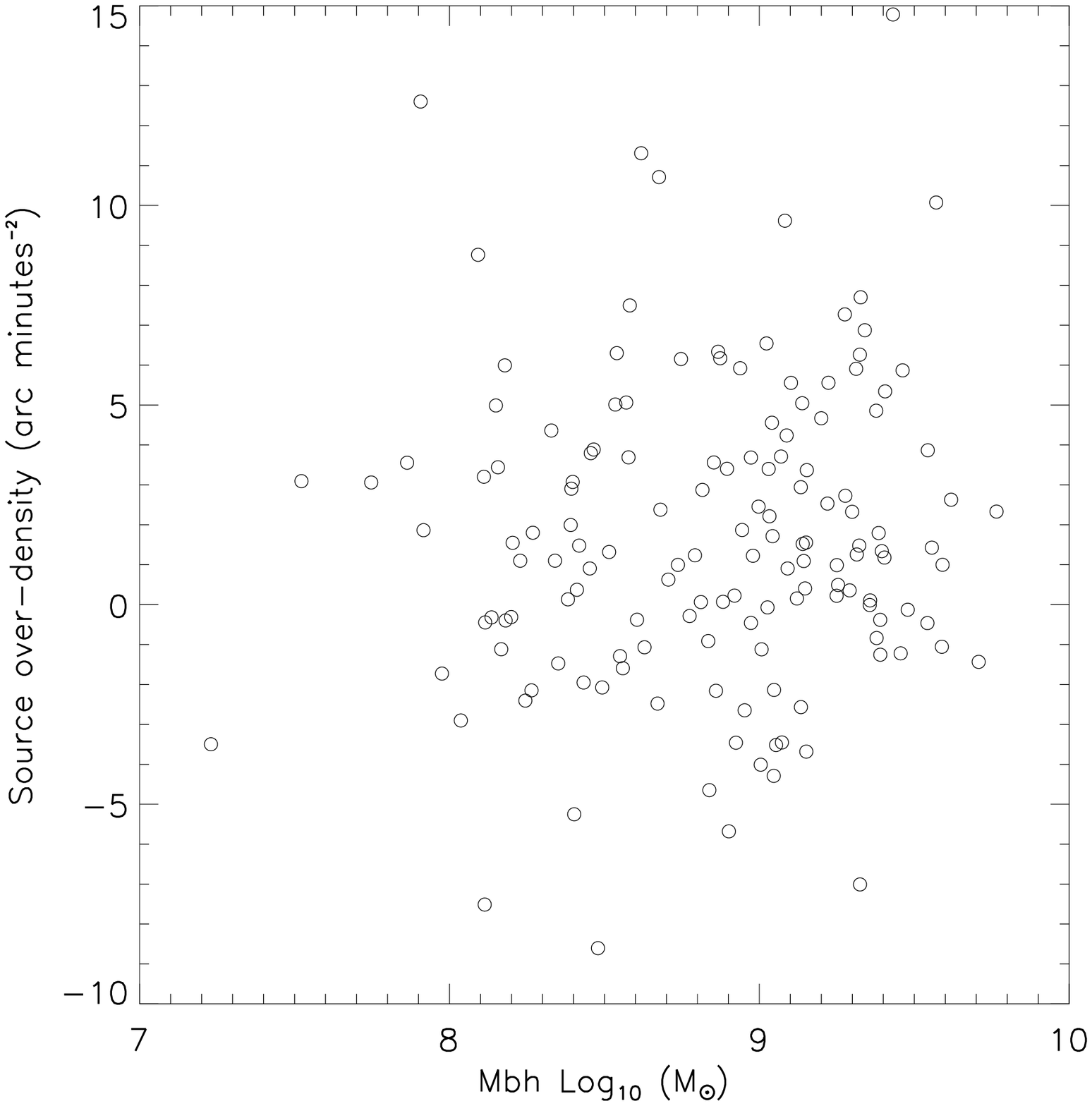}
\includegraphics[width=0.8\columnwidth]{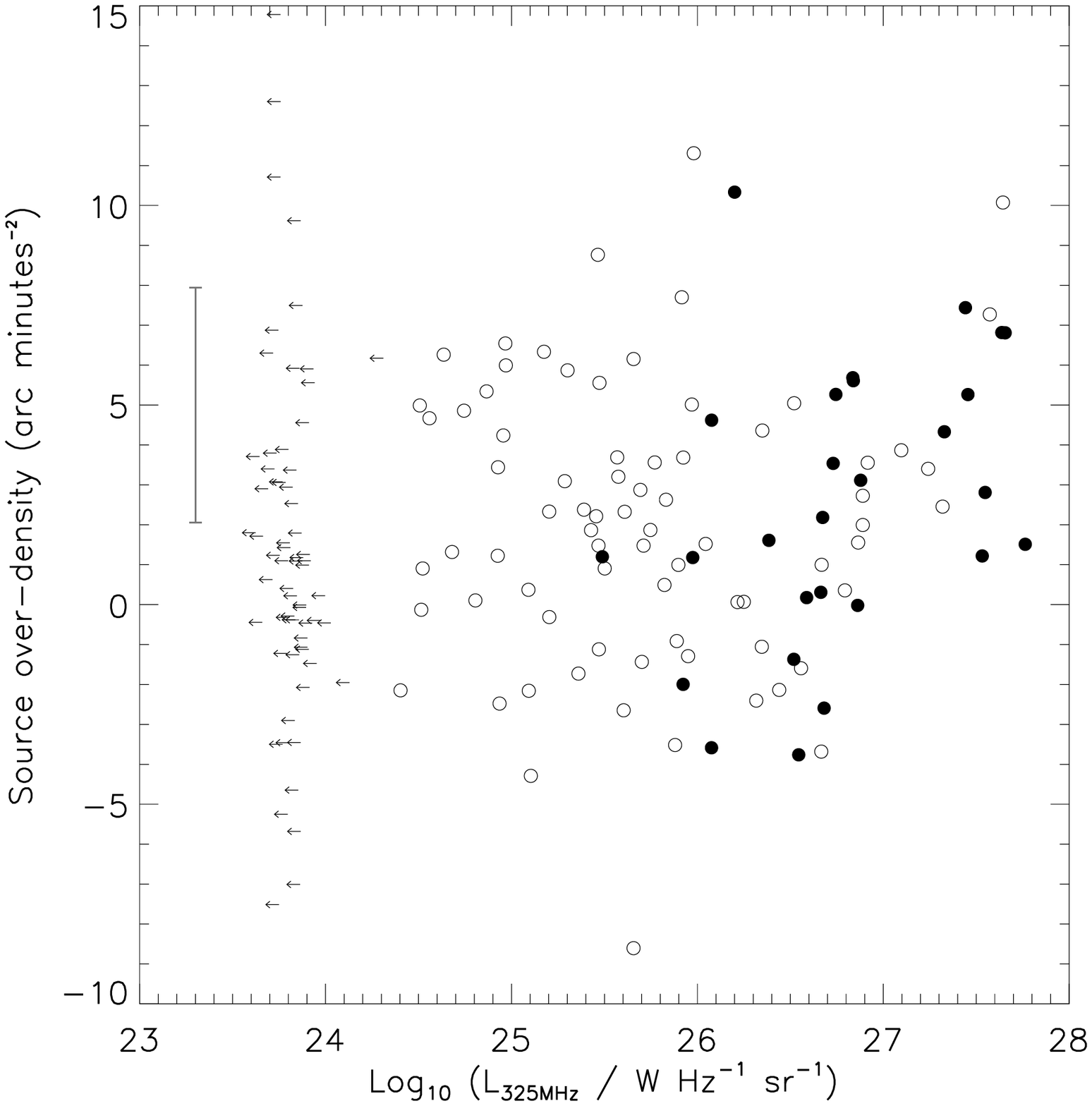}
\caption{Source density within $\sim$300~kpc at z$\sim$1 as a function
  of black-hole mass for the quasars (left-hand panel) and as a
  function of radio luminosity at $325$~MHz for all AGN (right-hand
  panel). In the right-hand panel, the open circles show the RLQs, the filled
  circles the RGs and the upper limits the RQQs; also shown in this
  panel is an example of the size of the error bars on the
  over-densities in both panels. The error bars would show the Poisson
  error for each AGN field combined in quadrature with the Poisson error on
  the blank field level, that was subtracted, both scaled by their
  mean completeness correction.}
\label{fig:correlation_density}
\end{figure*}

We can go one step further by testing whether over-density is related to
black-hole mass regardless of AGN classification. This is shown in the left
panel of Fig.~\ref{fig:correlation_density}. To search for possible trends we
performed a correlation analysis on the AGN over-density measurements in the
first bin. We have used two types of correlation analysis: the Spearman rank
order test and the Generalized Kendall's tau test. Both tests suggest a low
probabilty of a correlation (see Table~\ref{tab:correlation}).

A possible source of uncertainty in the black-hole masses of RQQs and RLQs
could be their orientations with respect to the observer. As shown by
\cite{Jarvis&McLure02} and \cite{Jarvis&McLure06} one would expect that sources
with bright core radio emission would be preferentially aligned pole-on to the
observer. Coupled with a disc-like broad-line region, such a bias would result
in lower derived black-hole masses for the RLQs relative to the RQQs given the
same optical selection. However, our initial selection, based on low-frequency
radio emission using the WENSS at 325~MHz, means that we minimize beaming
effects as the radio emission at low frequencies is dominated by the extended,
optically thin, lobe emission. Therefore, although we cannot rule out
completely a link between environmental density and black-hole mass for our
sample, it would seem very unlikely.

\begin{table*}
\centering
\caption{Correlation analysis on the data from
  Fig.~\ref{fig:correlation_density}, i.e. source density within $\sim$300~kpc
  at $z\sim$1 vs black-hole mass for the quasars and then vs radio luminosity
  for all AGN. AGN(L$_{\rm radio} >$ 25 or 26) corresponds to AGN with
  Log$_{10}(L_{325}/\rm{W\,Hz}^{-1}\,\rm{sr}^{-1}) >$ 25 or 26. Coefficients
  shown are the correlation coefficient except for the Cox hazard test which
  gives a $\chi^2$ value. The significance shown is the confidence level at
  which the null hypothesis (i.e. no correlation) can be rejected. Survival
  analysis was used on the radio luminosity data as these contain upper limits
  on the RQQs radio luminosity.}
\label{tab:correlation}
\begin{tabular}{cccc} 
\hline
 & Test & Coefficient & Significance \\
\hline  
Quasars & Spearman rank & 0.080 & 0.666 \\
(M$_{\rm BH}$ vs over-density)  & Kendall tau & 0.054 & 0.669 \\  
\hline
All AGN & Spearman rank& 0.160 & 0.961 \\
(radio luminosity vs over-density)& Kendall tau & 0.215 & 0.966 \\
 & Cox hazard & 2.109 & 0.854 \\
\hline
AGN(L$_{\rm radio}$$>$ 25) & Spearman rank & 0.226 & 0.963 \\
(radio luminosity vs over-density)& Kendall tau & 0.155 & 0.966 \\
\hline
AGN(L$_{\rm radio}$$>$ 26) & Spearman rank & 0.483 & 0.992 \\
(radio luminosity vs over-density)& Kendall tau & 0.121 & 0.999 \\
\hline
\end{tabular}
\end{table*}

\subsection{Radio luminosity vs environmental density}

Having upper limits for the RQQ radio power (see Section~\ref{section:sample})
allows us to investigate the environmental densities of all our AGN as a
function of radio luminosity (see Fig.~\ref{fig:correlation_density}). Since
the RGs are not selected in the same way as the quasars, caution must be
exercised in their inclusion in this analysis. It is also worth noting the size
of the error bars on the over-density in these figures, as shown by the mean
error bar ($\pm 3$ arcmin$^{-2}$). This explains the large scatter on any
correlation.

The censored data (e.g. the RQQ upper limits) require that we use survival
analysis techniques for our correlation analysis. We therefore used {\sc iraf}
which provides three types of test: the Spearman rank order test, the
Generalized Kendall's tau test and the Cox proportional hazard model. All three
tests can handle one type of limit in the dependent variable, in our case upper
limits in the radio luminosity. The results of the correlation analysis are
shown in Table~\ref{tab:correlation}. With the exception of the Cox
proportional hazard model the tests give evidence for a correlation at the
$>95$~per cent confidence level when all AGN are included. We also restrict our
analysis to those sources that have radio luminosities above the traditional
separation between FRI and FRII radio sources at
Log$_{10}(L_{325}/\rm{W\,Hz}^{-1}\,\rm{sr}^{-1} )= {25}$ (see
e.g. \citealt{CJ04}).  At this threshold there is also a divergence in the
space-density evolution with redshift, with the higher luminosity radio sources
tending to evolve more strongly than the lower luminosity sources
(\citealt{CJ04}; \citealt{Sad07}). For this sample we again obtain correlation
at the $>95$ per cent level. However, if we employ a much more conservative
low-luminosity cut-off of Log$_{10}(L_{325}/\rm{W\,Hz}^{-1}\,\rm{sr}^{-1}) =
26$ which ensures that all of our sources lie well within the FRII r\'egime
then we find that the correlation is significant at the $>99$ per cent level.

Intriguingly, in the high radio luminosity range the RLQs and RGs appear to
show the same trend of increasing source over-density with radio luminosity, as
would be expected in the unified scheme \citep{Barthel89}. This improvement in
the correlation analysis results for high radio luminosity could of course be a
real effect or instead it might be a result of foreground/background
contamination in some of the low-radio luminosity AGN causing them to appear
more over-dense than they actually are. However, recent work by \cite{Donoso09}
finds that the environmental densities of RLQs and radio-loud AGN (i.e. RGs)
match only for radio luminosities of
Log$_{10}(L_{325}/\rm{W\,Hz}^{-1}\,\rm{sr}^{-1}) \simgt 25.4$ (after converting
into appropriate units). Their interpretation is that the unified scheme for
radio-loud quasars and radio galaxies might only be valid for high radio
luminosities.  Finally, we note with interest that \cite{Donoso09} also report
a similar trend to that found here of increasing environmental density with
radio luminosity for RLQs.

\section{Discussion}

Our results indicate that, on average, AGN at $z\sim1$ have 2 to 3 massive
($\gtsim L^*$) galaxies containing a substantial evolved stellar population in
their $\sim300$~kpc-scale environments, representing an over-density relative
to the field.

Moreover, we find evidence that the radio emission we observe from AGN is in
some way related to the galaxy density in its environment. Specifically, we
find that the RLQs and RGs occupy more dense environments than the
RQQs. Secondly, if we ignore AGN classification and simply measure the
over-density as a function of radio luminosity
(Fig.~\ref{fig:correlation_density}) we find evidence of a positive correlation
between the two (Table~\ref{tab:correlation}).

Initially one might expect that these results could be due to the
radio-loud objects having intrinsically larger black hole masses than
their radio-quiet counterparts. Indeed, recent work has found evidence
for this to be the case (e.g. \citealt{Lacy01,McLure&Jarvis04}). If we
then assume that the largest black holes reside in the largest dark
matter haloes, or are found closer to the centre of their haloes, they
would thus also have the highest density environments. However, when
we compare our results to the quasar black hole mass estimates we find
that the black-hole masses of the RLQs and RQQs are statistically
indistinguishable. Moreover, the over-density around the quasars shows
no significant trend with black-hole mass
(Fig.~\ref{fig:correlation_density}) using a correlation analysis
(Table~\ref{tab:correlation}).

At this point it is worth considering whether there might be a systematic bias
in the way our black-hole mass estimates are made for the radio-loud and
radio-quiet subsets. For example, if the Eddington ratios are for some reason
systematically lower for the radio-loud objects then they must host more
massive black-holes to produce a given optical luminosity. One possibility is
that radio-loud objects are powered by a radiatively inefficient accretion
process such as Bondi accretion of the hot phase of the IGM
(e.g. \citealt{HEC07}) while the radio-quiet objects are accreting cold gas in
the standard manner. However, such a possibility is easily dismissed since
radiatively inefficient accretion processes can only explain the
multi-wavelength properties of low-excitation radio sources which are almost
all Fanaroff-Riley (FR) class I objects whereas our entire sample is made up of
QSOs, with the radio-loud objects all having radio luminosities typical of FR
II sources (Fig.~\ref{fig:radioL_z}). Furthermore, given that our samples are
matched in absolute optical magnitude and optical colours, any difference in
accretion properties must contrive to produce a distribution of \mg line widths
that would lead to identical black-hole mass distributions; this seems
unlikely.

Therefore we are led to conclude that the environments of the AGN are somehow
affecting the differences we observe in their radio properties. This is one of
two possible scenarios that could explain our results, the other being that the
AGNs radio emission is influencing the environmental density. However, this is
a much harder scenario to envisage as the $\sim$100 kpc-scale radio jets would
need to influence galaxy formation on Mpc scales.

It is known that tidal stripping would be more prevalent in denser
environments, as more close encounters or mergers with other galaxies would
occur. The inter-galactic medium (IGM) would therefore be denser in regions of
higher galaxy density. A higher IGM density gives more material for radio jets
to work on, increasing the radio luminosity produced through synchrotron
losses. This effect, known as jet confinement, was discussed by
\cite{Barthel&Arnaud96} who use it to explain the unusually steep far-infrared
to radio spectral slope of Cygnus A as boosting of the radio luminosity caused
by a higher environmental density. The estimated enhancement in radio
luminosity for AGN in clusters, compared to the field, was given by
\cite{Barthel&Arnaud96} as $\sim1.5$ orders of magnitude which might be
sufficient to explain our results, although a realistic physical model is
clearly required. In the local universe \cite{KauffmannHeckman&Best} find a
similar difference in matched samples of radio-loud and radio-quiet emission
line AGN from SDSS; they also offer an explanation in terms of radio jets being
enhanced in denser environments. Our results extend this relationship to higher
luminosity objects at higher redshifts.

The idea of jet confinement may be able to explain the results for the RGs and
RLQs. Whether this idea can be extended to the RQQs, which typically have radio
emission at least an order of magnitude lower than the RLQs, is less
clear. Classically in unified schemes, RQQs are thought of as the quasars
without kpc-scale radio jets, implying that they are physically different to
RLQs. However, there is some evidence in the literature to suggest that the
radio properties of RQQs and RLQs are not so very different (see
\citealt{Lacy01} for example). It has been suggested though, that the use of
the FIRST survey, which is not sensitive to extended radio emission, may in
fact have been responsible for these findings (\citealt{Ivezic}). However,
\cite{Kukula98} detect kpc-scale radio jets around 34~per cent of a sample of
27 RQQs. If this is representative of the whole RQQ population then perhaps the
classical view is no longer valid and there is infact more of a continuum of
quasar radio properties. This certainly makes the source of the radio emission
from AGN easier to understand, as explaining a dichotomy in their properties is
difficult.  In our sample, there is definitely a gap in the radio luminosities
of our quasars (see Fig.~\ref{fig:radioL_z}) in the sense that the RQQ upper
limits are typically an order of magnitude below the least radio-loud
RLQs. This effect is, however, intrinsic to our sample, since we have used
different surveys which have different depths to define the RLQ and RQQ
sub-samples. Therefore this is not suggestive of a radio power dichotomy in the
quasar population although at the same time it is not evidence that there is
not one.

An alternative explanation is that the differences we observe in AGN radio
properties are caused by the spin states of their black-holes. This is
certainly plausible as it could much more readily explain a radio power
dichotomy, if indeed it turns out that there is one. It could also explain our
results if for some reason the spin is affected by the environmental
density. This idea is mentioned in the literature and usually takes the form of
black holes spinning faster in dense environments due to increased exposure to
mergers, which spin up the black hole (e.g. \citealt{Wilson&Colbert95},
\citealt{Moderski98}, \citealt{Volonteri07} and \citealt{Sikora07}). However,
the one major problem with black hole spin as an explanation is that it is as
yet not observationally measurable in AGN and so the hypothesis is not yet
testable.

\section{Conclusions}

We have conducted an analysis of the environments of a large sample of
AGN at $z\sim1$ in order to study the relationship between AGN
activity and environmental density at an epoch close to the peak in
AGN activity.  Our main conclusion are:

\begin{enumerate}

\item The AGN fields show, on average, an excess above the field of 2 to 3
  massive galaxies containing a substantial evolved stellar population. Most of
  this over-density is confined within a radius of $200-300$~kpc of the AGN
  although there is good evidence for a lower level over-density extending out
  to the Mpc-scale.

\item We find evidence for a trend of increasing galaxy over-density with
  increasing AGN radio luminosity. Since the RLQs and RQQs have
  indistinguishable black hole mass distributions the observed difference in
  environmental density is not a result of observing different populations of
  objects. This leads us to conclude that the radio power of an AGN is in some
  way influenced by the environmental density in which it resides.

\item Our results could be explained by the boosting of radio jets in areas of
  higher IGM density which are known to exist in galaxy-dense regions due to
  mergers and tidal stripping of galaxy gas. It is unclear whether this
  explanation can be extended to the RQQs in our sample.

\item It is of course entirely possible that the radio properties of AGN are
  not determined by a single parameter but instead a combination of
  parameters. It is clear from previous work that for an AGN to be radio-loud
  it requires a certain mass of black hole (\citealt{McLure&Jarvis04}) but this
  cannot be the only factor involved. This work and that of others
  (e.g. \citealt{KauffmannHeckman&Best} and \citealt{Donoso09}) suggests that
  there is some link between radio-loudness and the environment as well, and
  there have been several theoretical papers proposing that black hole spin
  could also be responsible.

\end{enumerate}

Future observations, such as deep radio observations with the Low-Frequency
Array (LOFAR), will allow us to investigate the environments as a function of
radio luminosity well down into the radio-quiet quasar r\'egime and up to
higher redshifts, and place firmer constraints on our conclusions based on this
sample alone.

\section*{ACKNOWLEDGMENTS}

This work is based [in part] on observations made with the {\em Spitzer Space
  Telescope}, which is operated by the Jet Propulsion Laboratory, California
Institute of Technology under a contract with NASA. Support for this work was
provided by NASA through an award issued by JPL/Caltech. JTF thanks the Science
and Technology Facilities Council for a research studentship. MJH thanks the
Royal Society for a research fellowship. RJM acknowledges funding from the
Royal Society. This research has made use of the NASA/IPAC Extragalactic
Database (NED) which is operated by the Jet Propulsion Laboratory, California
Institute of Technology, under contract with the National Aeronautics and Space
Administration.

\end{document}